\documentclass{article}

\usepackage[T1]{fontenc}
\usepackage{spconf,amsmath,graphicx,booktabs,array,cite,url}
\usepackage[hidelinks]{hyperref}
\usepackage{balance}

\title{UNHEARD BUT RECOGNIZABLE: ULTRASONIC SIGNATURES FOR MUSICAL\\
INSTRUMENT RECOGNITION}

\name{Izhak Kapash \qquad Uri Rom \qquad Ram Zamir}
\address{Tel Aviv University\\
\texttt{izhakkapash@mail.tau.ac.il}\\[-0.2em]
\texttt{urom@tauex.tau.ac.il} \quad
\texttt{zamir@eng.tau.ac.il}}

\begin{document}
\ninept
\setlength{\emergencystretch}{2em}
\maketitle

\begin{abstract}
Musical-instrument recognition is a fundamental music information retrieval (MIR) task supporting transcription, source separation, and audio understanding. Despite significant algorithmic progress, there is still room for improvement in recognizing acoustically similar instruments, particularly in polyphonic mixtures. We examine the contribution of ultrasonic components, i.e., frequencies above the audible 0--20 kHz range. Although most of this content cannot be represented at conventional 44.1/48-kSPS sampling rates under the Nyquist criterion, it is available in high-quality 96-kSPS recordings. Using a 96-kSPS multitrack corpus covering 15 classes across performers, instruments, studios, and sessions, we compare full-band, low-pass, and ultrasonic-only inputs using classical machine-learning and deep-learning classifiers. Our results show that source-dependent ultrasonic extensions and broadband transients contain instrument-specific information that significantly improves isolated and polyphonic recognition and motivates wider-bandwidth studies of other MIR tasks.
\end{abstract}


\section{Introduction}
\label{sec:introduction}

Musical-instrument recognition is a fundamental music information retrieval (MIR) task supporting transcription, indexing, source separation, and polyphonic audio analysis \cite{herrera2003instrument,muller2021fmp}.
Although modern signal-processing and learning systems perform well on broad instrument categories, distinguishing sources with similar audible-band spectra remains difficult, especially when several instruments are mixed together. Most MIR datasets and systems use recordings sampled at 44.1 or 48 kSPS, as exemplified by MedleyDB and MUSDB18 \cite{bittner2014medleydb,rafii2017musdb18}. These sampling rates impose Nyquist limits of 22.05 and 24 kHz, respectively \cite{oppenheim2010dsp}; accordingly, information above those frequencies is removed before a recognition model can assess its usefulness.

From a physical perspective, ultrasonic content in musical signals arises mainly through three mechanisms: extension of the harmonic series beyond 20 kHz, transient excitation generating broadband high-frequency energy, and noise-like radiation, particularly from metallic percussion \cite{boyk1997life,petrosino2016ultrasonic}. These mechanisms produce structured, source-dependent ultrasonic signatures in many acoustic instruments and vocal signals, while the added band may remain weak or absent in sources such as piano and clean magnetic-pickup electric guitar. Our preliminary guitar-only study suggested that these components can improve guitar-type recognition \cite{kapash2026guitar}; the present study extends this investigation to 15 source classes, multiple recording settings, polyphonic mixtures, and matched bandwidth controls. Higher sampling rates preserve these components and have long been used in professional recording workflows \cite{dunn1998filters,craven2004antialias,melchior2019history}. However, previous physical measurements did not test whether the added band improves learned instrument recognition under controlled, bandwidth-matched conditions.

We ask whether spectral information routinely discarded as inaudible can help identify instrumental and vocal sources. Raising the sampling rate changes both sample count and retained bandwidth; therefore, a direct 44.1-versus-96-kSPS comparison cannot attribute improved performance to ultrasonic content rather than sampling rate. We use matched 96-kSPS low-pass and ultrasonic-only controls. Our contribution is threefold: {\em (i)} a native 96-kSPS multitrack corpus of instrumental and vocal sources spanning multiple studios, performers, and physical instrument instances; {\em (ii)} controlled full-band, low-pass, and ultrasonic-only experiments; and {\em (iii)} evidence from both 15-class isolated-source classification and practical multi-label detection in polyphonic music. The aim is to test the informational value of added bandwidth above 22.05 kHz rather than to rank recognition architectures.

Compared with 44.1-kSPS input, 96-kSPS full-band input increased convolutional neural network (CNN) accuracy from 87\% to 91\% and random-forest macro-F1 from 81.11\% to 88.38\%. Polyphonic accuracy increased by approximately four percentage points, while ultrasonic-only input in the 20--48-kHz band achieved 82\% accuracy across 12 classes.


\section{Method}
\label{sec:method}

Recordings used Earthworks QTC50 and QTC30 omnidirectional microphones, specified by the manufacturer as 3 Hz--50 kHz and 3 Hz--30 kHz, respectively, with (+1/-3)-dB tolerances \cite{earthworks_qtc50,earthworks_qtc30}. Professional studio microphones generally do not impose an active 20-kHz cutoff; our end-to-end chain imposed none. A band-limited front end cannot benefit from 96-kSPS ultrasound. Signals passed through a Soundcraft console to a MOTU 2408 MK3 analog-to-digital (A/D) converter and a Universal Audio Apollo Twin interface, and were recorded natively in Cubase 10 at 96 kSPS (48-kHz Nyquist limit).

Each source was close-miked separately under acoustically isolated conditions, without effects or dynamic processing; aligned mixtures used only minor level balancing. Different sessions used different performers and different physical instrument instances within the shared family labels; studios and microphone configurations also varied. A complete QTC50-only session reproduced the same relative ultrasonic patterns. We treat the measurements as properties of the acquired signals rather than calibrated free-field source spectra.

All conditions were derived digitally from the same native 96-kSPS takes: matched 44.1-kSPS signals, a 96-kSPS low-pass (LP) control at 22.05 kHz, and an ultrasonic-only control retaining 20--48 kHz. Thus, microphone and preamplifier response are identical within each paired bandwidth comparison. Segment durations, recording-level splits, model settings, and analysis windows were matched. Implementation details and resources are public \cite{kapash2026resources}.


\section{Dataset}
\label{sec:dataset}

The corpus contains six complete multitrack pieces---two jazz, two blues/rock-and-roll, one Andalusian, and one minor-key ballad---totaling over one hour of mixtures and nearly 15 hours of aligned isolated sources. Although modest in repertoire size, the data span multiple studios, sessions, performers, microphone configurations, and physical instruments across the 15 classes in Table~\ref{tab:ultrasonic}. Several labels therefore represent instrument families rather than single exemplars. Class-balanced window sampling reduces imbalance caused by unequal stem durations.

Splitting preceded segmentation. All CNN and random-forest results use recording-disjoint partitions: complete takes were assigned before window extraction, and every class had distinct training and evaluation takes. The same partitions were used across bandwidth conditions. Training ran for a fixed 50 epochs; held-out takes did not select checkpoints or thresholds. Results are three-seed means with standard deviation (SD) at most one percentage point. These statistics characterize repeatability, not population significance; the principal four-to-six-point gains exceed run variation, whereas the 1.15-point low-pass--44.1 difference is minor. Recordings, splits, code, and reproduction instructions are public \cite{kapash2026resources}.


\section{Ultrasonic Characteristics}
\label{sec:ultrasonic}

\begin{figure*}[t]
  \centering
  \begin{minipage}[t]{0.44\textwidth}
    \centering
    \includegraphics[width=\linewidth,trim=0 0 0 65,clip]{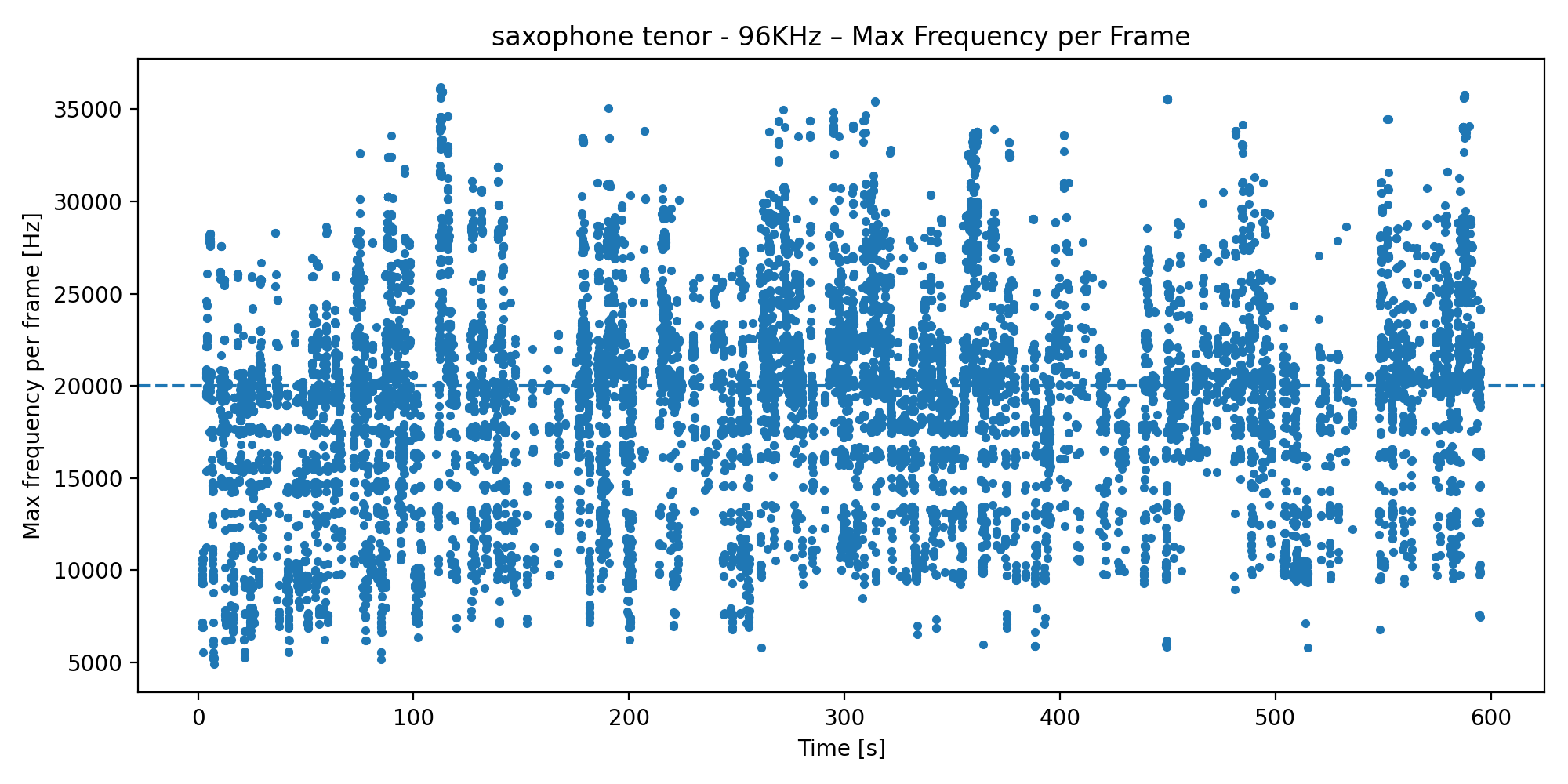}\\[-1mm]
    (a) Tenor saxophone
  \end{minipage}\hfill
  \begin{minipage}[t]{0.44\textwidth}
    \centering
    \includegraphics[width=\linewidth,trim=0 0 0 65,clip]{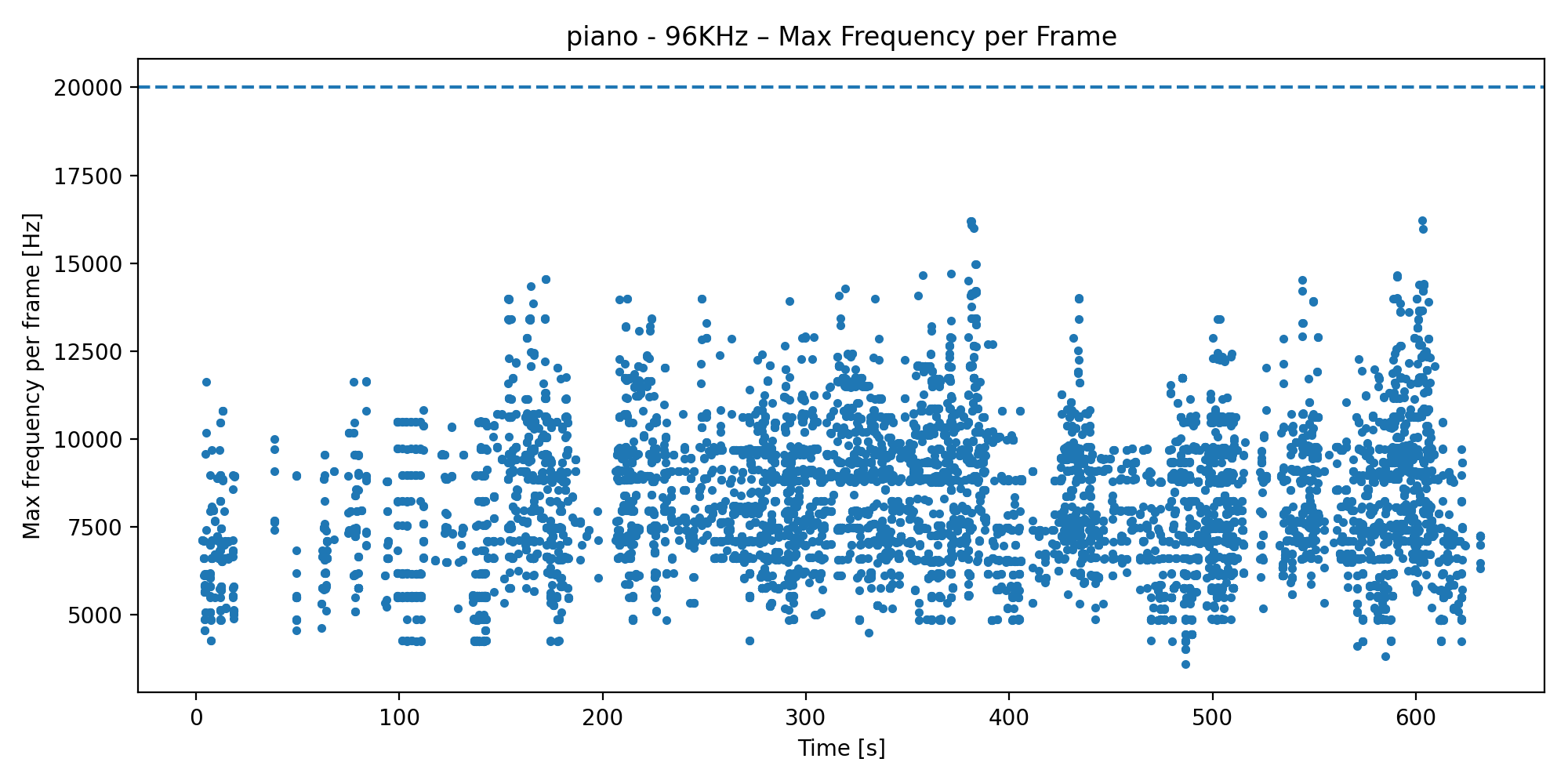}\\[-1mm]
    (b) Piano
  \end{minipage}
  \caption{Frame-wise maximum detected frequency at 96 kSPS. Activity is shown only 50 dB above the estimated noise floor; this threshold is descriptive and is not used by the recognition models. Dashed lines mark 20 kHz.}
  \label{fig:contrast}
\end{figure*}

For descriptive analysis, a short-time Fourier transform (STFT) yielded the highest active frequency per frame for each isolated track. Components were counted only when at least 50 dB above the estimated studio noise floor. This pragmatic threshold aids display in Fig.~\ref{fig:contrast} and the PUA summary in Table~\ref{tab:ultrasonic}; it is not a physical boundary and is never used for model input or labels. A frame maximum is sensitive to isolated peaks and therefore describes spectral extent only; prevalence is measured separately by activity probability.

\begin{figure}[t]
  \centering
  \includegraphics[width=0.94\linewidth,trim=0 0 0 38,clip]{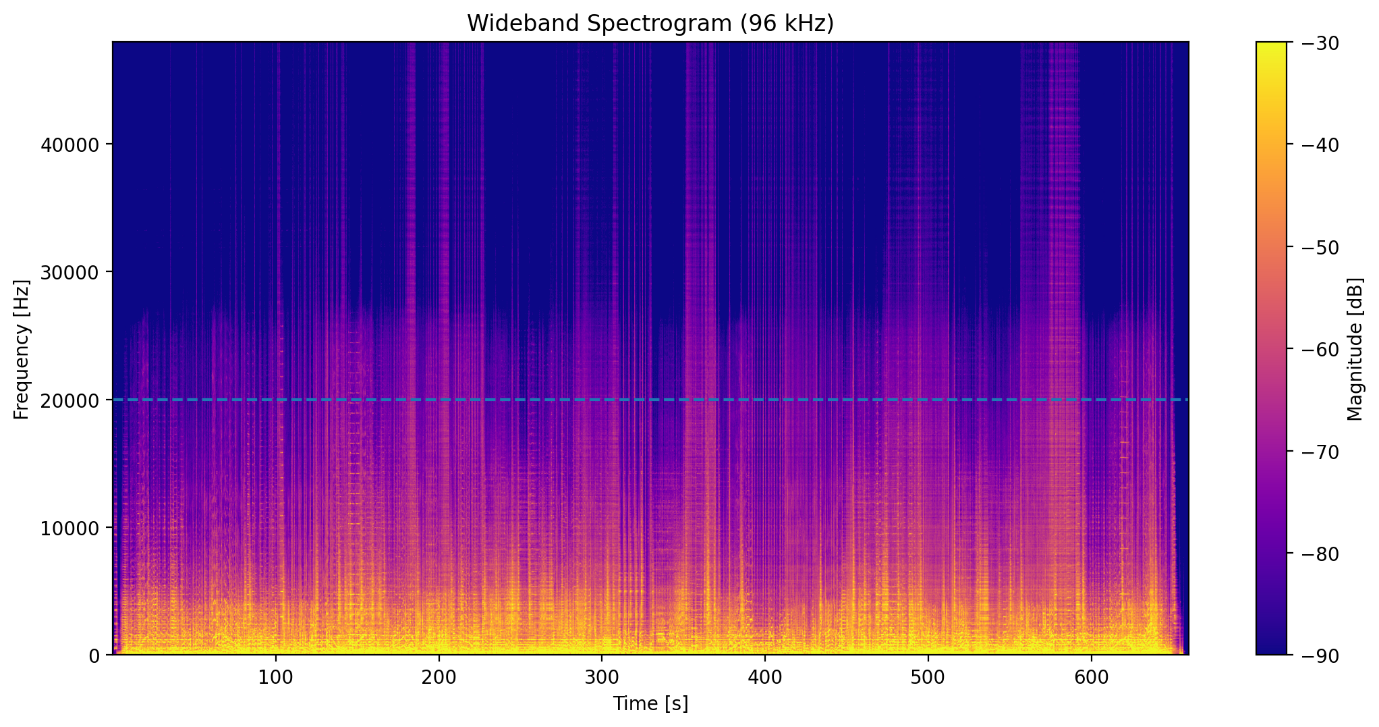}
  \caption{Wideband spectrogram of a 96-kSPS polyphonic ensemble recording. The dashed line marks 20 kHz; spectral structure and broadband transients extend above it.}
  \label{fig:wideband}
\end{figure}

The added band exhibits distinct morphologies. Sustained sources form stable low-ultrasonic trajectories consistent with higher-order partials, whereas plucked-string and percussion attacks produce short broadband structures that can reach the 48-kHz ceiling; airflow and metallic excitation may yield diffuse noise-like energy \cite{boyk1997life,petrosino2016ultrasonic}. Thus, persistence and bandwidth provide structure beyond total high-frequency energy. In Fig.~\ref{fig:contrast}, saxophone repeatedly extends beyond 30 kHz, whereas piano shows no meaningful ultrasound. Clean electric guitar behaves similarly, consistent with magnetic-pickup low-pass behavior \cite{paiva2012pickups}; acoustic and classical guitars retain ultrasonic extensions.

\begin{table}[t]
\centering
\caption{Pooled ultrasonic activity above threshold. The probability of ultrasonic activity (PUA) is the proportion of one-second intervals above threshold. An asterisk marks the 48-kHz measurement ceiling.}
\label{tab:ultrasonic}
\setlength{\tabcolsep}{3.2pt}
\renewcommand{\arraystretch}{0.98}
\begin{tabular}{@{}lcc@{}}
\toprule
Instrument & PUA (1 s) & Highest detected (kHz) \\
\midrule
Bansuri          & 0.21 & 30.4 \\
Classical guitar & 0.44 & 48.0$^\ast$ \\
Double bass      & 0.69 & 38.0 \\
Flute            & 0.49 & 28.2 \\
Kanun            & 0.92 & 48.0$^\ast$ \\
Tenor saxophone  & 0.81 & 36.2 \\
Trumpet          & 0.34 & 27.4 \\
Violin           & 0.89 & 27.9 \\
Vocals           & 0.80 & 26.5 \\
Acoustic guitar  & 0.94 & 28.1 \\
Clarinet         & 0.61 & 26.8 \\
Electric guitar  & 0.00 & 17.1 \\
Harmonica        & 0.99 & 27.7 \\
Piano            & 0.00 & 16.2 \\
Drums            & 1.00 & 48.0$^\ast$ \\
\bottomrule
\end{tabular}
\end{table}

PUA is the fraction of valid one-second intervals containing above-threshold energy above 20 kHz, not an instrument constant. Dynamics, articulation, performer, and exemplar affect this probability; stronger excitation can yield more high-frequency events. PUA often correlates with recognition but does not predict it: exceptions show that models exploit full class-dependent distributions rather than fixed spectral cutoffs. Highest detected frequency instead describes spectral extent, so sparse diagnostic events can coexist with low PUA. Maxima characterize the acquired signal rather than an absolute source limit, and 48-kHz values are right-censored by Nyquist.


\section{Classification and Recognition Tasks}
\label{sec:tasks}

This section presents three experiments assessing the value of the added ultrasonic band: 15-class isolated-source classification, multi-label ON/OFF detection for selected sources in annotated polyphonic excerpts, and ultrasonic-only classification over a fixed 12-class subset. All tasks use the recording-level protocol described in Section~\ref{sec:dataset}.

\subsection{Isolated-source classification}
\label{sec:isolated}

Because the objective is to isolate the effect of extending spectral bandwidth rather than to benchmark architectures, we evaluated two complementary classifiers: CNN6, a convolutional neural network, and a feature-based random forest. Consistent results across these substantially different paradigms make it less likely that the observed effect depends on one particular representation or learning mechanism.

CNN6 was adapted from Pretrained Audio Neural Networks (PANNs) \cite{kong2020panns,gemmeke2017audioset}; related embeddings have been evaluated for instrument classification \cite{ding2023embeddings}. Inputs were per-bin-normalized log-power linear short-time Fourier transforms (STFTs; 32-ms Hann, 8-ms hop; 3072 points/1537 bins at 96 kSPS). Four 5$\times$5 blocks feed frequency averaging, mean-plus-max temporal pooling, and a 512-unit embedding. Training from scratch avoided pretraining effects. Mean held-out accuracy was approximately 91\% at full-band 96 kSPS and 87\% at 44.1 kSPS. Low-pass filtering 96-kSPS input at 22.05 kHz returned accuracy to the 44.1-kSPS level, attributing the gain to retained bandwidth rather than sample count.

\begin{figure}[t]
  \centering
  \includegraphics[width=0.94\linewidth]{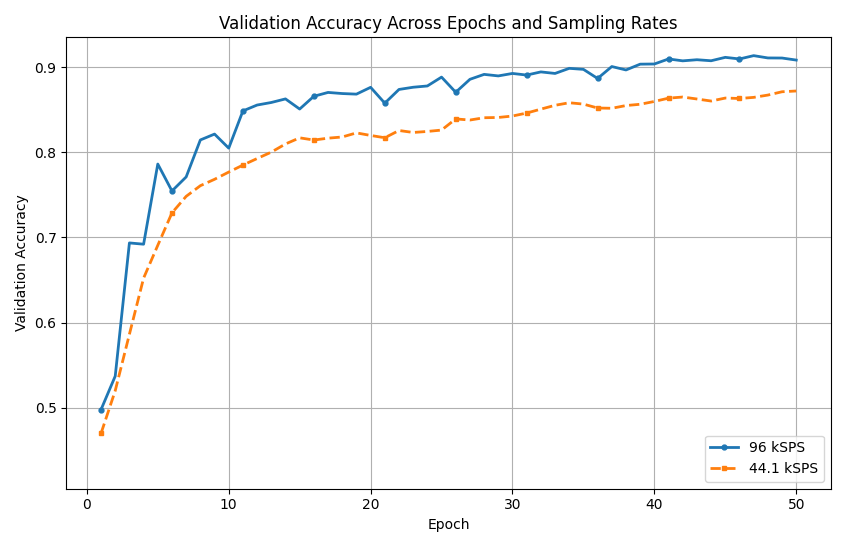}
  \caption{Mean validation accuracy over three runs. Full-band 96 kSPS remains above 44.1 kSPS. The near-overlapping 96-kSPS low-pass control is omitted for clarity. Final-accuracy standard deviations are at most one percentage point.}
  \label{fig:validation}
\end{figure}

The random forest \cite{breiman2001random} used paired active 0.5-s windows in every condition: up to 300 training and 150 evaluation windows per class from recording-disjoint takes. A linear STFT used 32-ms Hann windows and an 8-ms hop. Relative energy in fixed 2-kHz bands to 40 kHz balanced audible and ultrasonic capacity; centroid, bandwidth, roll-off, flatness, crest, slope, flux, and temporal statistics were added. We avoided mel-frequency cepstral coefficients (MFCCs), whose compression favors the audible range. Table~\ref{tab:rf} reports mean F1 over three seeds for a 200-tree forest; macro-F1 is the unweighted class mean.

\begin{table}[t]
\centering
\caption{Mean random-forest F1 over three runs; SD is at most one percentage point. LP denotes low-pass filtering at 22.05 kHz.}
\label{tab:rf}
\setlength{\tabcolsep}{2.4pt}
\renewcommand{\arraystretch}{0.98}
\begin{tabular*}{\columnwidth}{@{\extracolsep{\fill}}lccc@{}}
\toprule
Instrument & 96 kSPS & 44.1 kSPS & 96 kSPS LP \\
\midrule
Acoustic guitar  & 80.39\% & 77.50\% & 76.35\% \\
Bansuri          & 94.82\% & 80.38\% & 80.59\% \\
Clarinet         & 76.77\% & 75.72\% & 76.64\% \\
Classical guitar & 71.84\% & 63.31\% & 68.21\% \\
Drums            & 88.39\% & 80.40\% & 79.65\% \\
Double bass      & 94.45\% & 88.64\% & 88.96\% \\
Electric guitar  & 94.44\% & 82.31\% & 90.95\% \\
Flute            & 79.17\% & 81.09\% & 80.52\% \\
Harmonica        & 93.78\% & 90.29\% & 90.29\% \\
Kanun            & 93.79\% & 87.59\% & 88.00\% \\
Piano            & 93.58\% & 87.41\% & 89.55\% \\
Saxophone        & 89.99\% & 79.82\% & 80.14\% \\
Trumpet          & 93.29\% & 83.33\% & 83.41\% \\
Violin           & 92.24\% & 88.72\% & 88.98\% \\
Vocals           & 88.75\% & 70.18\% & 71.61\% \\
\midrule
Overall macro-F1 & 88.38\% & 81.11\% & 82.26\% \\
\bottomrule
\end{tabular*}
\end{table}

Macro-F1 was 88.38\% for full-band 96 kSPS, 82.26\% for 96-kSPS low-pass, and 81.11\% at 44.1 kSPS. The small 1.15-point low-pass--44.1 difference may reflect resampling, filtering, and finite-run variation; the 6.12-point full-band--low-pass gap instead identifies retained ultrasound as the principal source of improvement.

Large full-band gains over low-pass occurred for vocals (+17.1 F1 points), trumpet and saxophone (+9.9 each), drums (+8.7), and kanun (+5.79). Balanced sampling reduces class-count imbalance as an explanation. Electric-guitar and piano gains do not imply ultrasonic emission: weak or absent ultrasonic energy can itself be discriminative, while better separation of ultrasonic-active sources reduces false assignments to less-ultrasonic classes. Consistently, piano showed no material gain in the independent binary detectors of Section~\ref{sec:polyphonic}.

Using one feature-based model and one established CNN tests the effect across hand-engineered and learned representations. Recurrent and transformer architectures \cite{choi2017crnn,gong2021ast} remain future bandwidth-controlled tests rather than additional benchmarks here.

\begin{table*}[t]
\centering
\caption{Mean polyphonic results over three runs for selected sources; SD is at most one point. For each metric, 96 kSPS precedes 44.1 kSPS; entries are percentages.}
\label{tab:polyphonic}
\setlength{\tabcolsep}{3.0pt}
\renewcommand{\arraystretch}{0.96}
\begin{tabular*}{\textwidth}{@{\extracolsep{\fill}}lcccccccc@{}}
\toprule
& \multicolumn{2}{c}{Accuracy} & \multicolumn{2}{c}{Precision} & \multicolumn{2}{c}{Recall} & \multicolumn{2}{c}{F1} \\
\cmidrule(lr){2-3}\cmidrule(lr){4-5}\cmidrule(lr){6-7}\cmidrule(lr){8-9}
Instrument & 96 & 44.1 & 96 & 44.1 & 96 & 44.1 & 96 & 44.1 \\
\midrule
Drums            & 99.03 & 98.96 & 99.71 & 99.61 & 99.03 & 99.03 & 99.36 & 99.32 \\
Double bass      & 80.97 & 78.66 & 88.78 & 87.64 & 86.02 & 83.98 & 87.38 & 85.77 \\
Acoustic guitar  & 82.23 & 78.29 & 74.87 & 79.25 & 82.13 & 60.27 & 78.33 & 68.47 \\
Classical guitar & 88.55 & 74.42 & 92.74 & 68.02 & 84.87 & 96.89 & 88.63 & 79.93 \\
Tenor saxophone  & 83.50 & 74.80 & 81.40 & 80.32 & 84.20 & 73.46 & 82.78 & 76.73 \\
Harmonica        & 81.65 & 78.16 & 91.87 & 90.30 & 81.27 & 77.48 & 86.24 & 83.40 \\
Vocals           & 85.00 & 80.40 & 82.90 & 80.50 & 88.80 & 83.60 & 85.75 & 82.02 \\
\bottomrule
\end{tabular*}
\end{table*}


\subsection{Detection in polyphonic mixtures}
\label{sec:polyphonic}

Unlike the controlled attribution in Section~\ref{sec:isolated}, this task asks whether native 96-kSPS input improves practical polyphonic detection over 44.1 kSPS. CNN6 produced an ON/OFF decision for every target and segment. Across the complete target set, not only the rows in Table~\ref{tab:polyphonic}, 96-kSPS input improved overall multi-label accuracy by approximately four absolute percentage points. Accuracy counts correct decisions, precision predicted-activity reliability, recall detected activity, and F1 is their harmonic mean. Both rates used identical annotations and rules; the table reports representative percussion, string, wind, and vocal sources.

The effect is source dependent. Drums are saturated above 99\%. Classical-guitar precision rises from 68.02\% to 92.74\% despite lower recall, while acoustic-guitar recall rises from 60.27\% to 82.13\%; this agrees with acoustic guitars retaining ultrasound while clean electric guitar does not. Saxophone gains 8.70 accuracy points and vocals 4.60. Matched 96-kSPS low-pass aggregate performance was nearly identical to 44.1 kSPS and is omitted only for legibility. Thus, denser sampling alone contributes little; the material gain requires the retained band.


\subsection{Ultrasonic-only classification}
\label{sec:ultrasonic-only}

We removed content at and below 20 kHz and trained CNN6 on 20--48 kHz over 12 balanced classes (11 instruments and vocals). The cutoff follows the conventional hearing boundary, not the 22.05-kHz Nyquist limit: 20--22.05 kHz is technically representable at 44.1 kSPS, while Section~\ref{sec:isolated}'s low-pass control attributes the main 96-kSPS gain to added bandwidth. Evaluation used a complete held-out take. Mean accuracy over three runs was approximately 82\% with SD at most one point, versus 8.33\% uniform chance. Drums, kanun, and saxophone exceeded 90\% classwise F1, whereas flute and clarinet were near 70\%.

As an artifact control, we applied the same protocol to balanced quiet windows from unedited raw recordings, excluding digital zeros. Windows were retained only when full-band (0--48 kHz) root-mean-square (RMS) level was below -80 decibels relative to full scale (dBFS), before presenting their 20--48 kHz content. With the same recording-level split, performance was approximately at chance level and had no stable class pattern, making a simple stationary studio or chain signature unlikely. Thus, isolated instruments and voice remained identifiable after audible content was removed, motivating future inaudible polyphonic detection.


\balance
\section{Discussion and Conclusion}
\label{sec:discussion}

This work contributes from four perspectives. First, across 15 classes and diverse performers, exemplars, and sessions, most acoustic instruments exhibited structured ultrasound, extending earlier single-source measurements \cite{boyk1997life}. Vocals retained discriminative content; piano and clean magnetic-pickup electric guitar showed negligible activity. The sample spans common Western and Middle Eastern instrumentation.

Second, ultrasound improved a feature-based random forest (RF) and convolutional neural network (CNN), in isolated and polyphonic audio. Drums, kanun, and saxophone combined strong activity with high ultrasonic-only scores; flute and clarinet were nearer 70\%. Absence also separates electric from acoustic and classical guitars.

Across three matched runs, the main gains---four CNN accuracy points, 6.12 RF macro-F1 points over low-pass, and approximately four polyphonic-detection points---exceeded the at-most-one-point standard deviation (SD). With \(n=3\), this indicates repeatability, not population significance. Matched low-pass results followed 44.1 kSPS, so denser sampling contributed little. The 82\% ultrasonic-only result and chance-level quiet control argue against a stationary chain artifact.

Third, recognition without audible content demonstrates inaudible machine listening and suggests passive detection and localization. A strictly preliminary NMF case study on one percussion excerpt suggests improved clustering \cite{lee2001nmf,smaragdis2003nmf,ng2001spectral}: at 96 versus 44.1 kSPS, SDR changed from -4.07 to 2.15 dB, SI-SDR from -10.20 to -1.42 dB, and SIR reached 14.7 dB \cite{vincent2006bss,leroux2019sdr}. This does not establish general separation performance \cite{hennequin2020spleeter,defossez2021hybrid,rouard2023htdemucs}; Oohashi et al. offer a separate sensory perspective \cite{oohashi2000hypersonic}.

Fourth, 96-kSPS acquisition offers more than gentler filter transitions \cite{dunn1998filters,craven2004antialias,melchior2019history}: a wideband chain preserves information to 48 kHz. It requires 2.18 times the 44.1-kSPS storage, now less restrictive with modern hardware. Conventional copies can be made from a wideband master, whereas filtered ultrasound cannot be restored. We recommend native 96-kSPS masters through wideband front ends for future MIR, acoustic, or archival analysis \cite{iasa2009tc04}.

The main limitation is six compositions, despite over one hour of mixtures and nearly 15 hours of stems. Recording-disjoint evaluation tests beyond individual takes, while the corpus varies performers, physical instances, studios, and sessions; population generalization remains open. Larger corpora, per-recording PUA ranges, and broader separation tests are needed. Within this scope, the controls provide evidence that physically present musical and vocal ultrasound is computationally useful.

\section{Acknowledgments}

The authors thank Sharon Gannot for guidance, Harel Hadad and Sharon Reiz for recording, and all studio musicians.

\clearpage

\noindent\textbf{Funding:} The Samueli family donation for music-tech research.

\section{Compliance with Ethical Standards}

This musical-audio study did not involve human or animal subjects and required no ethics approval; the authors report no conflicts. \textbf{AI disclosure:} OpenAI ChatGPT and Anthropic Claude Code assisted with editing, code, and formatting; the authors verified all outputs and retain full responsibility.

\bibliographystyle{IEEEbib}
\bibliography{refs_revised}

\end{document}